\documentstyle[epsf]{l-aa}
\newcommand{\eqb}{\begin{eqnarray}}
\newcommand{\eqe}{\end{eqnarray}}
\newcommand{\mdot}{\stackrel{.}{m}}
\newcommand{\mcdot}{\stackrel{.}{M}}

\begin{document}
\thesaurus{02.01.2; 02.02.1; 02.18.5; 02.18.7; 08.14.1; 13.25.3}
\title{Spherical~accretion~onto~neutron~stars~and~black~holes}
\author{ L. Titarchuk\inst{1} A. Mastichiadis\inst{2},  
and N. D. Kylafis\inst{3}}
\institute{
NASA/GSFC and George Mason University/CSI
\and
Max-Planck-Institut f\"ur Kernphysik, Postfach 10 39 80,
D-69029 Heidelberg, Germany 
\and
University of Crete and Foundation for Research and Technology
- Hellas}
\offprints{L. Titarchuk}
\date{Received; accepted \dots}
\maketitle
\begin{abstract}
Spectral formation in steady state,
spherical accretion onto neutron stars and black holes is examined
by solving numerically and analytically the equation of radiative transfer.
The photons escape diffusively and their energy gains come
from their scattering off thermal electrons in the converging flow
of the accreting gas.
We show that the bulk motion of the flow
is  more efficient in upscattering photons
than thermal Comptonization in the range of non-relativistic electron
temperatures.
The spectrum observed at infinity is a power law with an exponential 
turnover at energies  of order the electron rest mass.
Especially in the case of accretion into a black hole,
the spectral energy power-law index is distributed around 1.5.
Because bulk motion near the horizon (1-5 Schwarzschild radii)
is most likely  a necessary characteristic of accretion into a black
hole, we claim that {\it observations of an extended power law
up to about $m_ec^2$, formed as a result of bulk motion Comptonization,
is a real observational evidence for the existence of an underlying 
black hole}.
\end{abstract}

\section{Introduction}
Spherical accretion into black holes as a Compton upscattering problem
of low frequency photons was 
studied by Blandford and Payne
(Blandford \& Payne 1981a, henceforth BP; 
Payne \& Blandford 1981, henceforth PB).
Especially in PB the above authors solved the photon transfer equation in the
case of a steady state spherically symmetric supercritical accretion
into a central black hole with the assumptions of power law flow
velocity and cold electrons $T_e=0$.
An extension of the work of PB
on accreting black holes to include the thermal 
motions of the electrons along with the bulk motion was performed 
by Colpi (1988). However, in the papers mentioned above, the authors
solved the photon transfer equation under the assumption 
that the inflow extends up to the center ($r=0$) and thus 
its thickness is infinite. 

Lyubarskij \& Sunyaev (1982)
solved the radiation transfer equation in the case of a {\it finite} optically 
thick medium. In a extension of the work of Blandford \& Payne (1981b), 
they studied Comptonization in a radiation-dominated shock, 
taking into account not only the bulk motion of the electrons but also
their thermal motion. 

In this paper we present the emergent 
spectrum from a spherical inflow by taking into account the effects of
bulk motion and thermal Comptonization in addition to the boundary conditions
of the inflow. For proofs and detailed discussions the reader is referred
to our long paper
(Titarchuk, Mastichiadis and Kylafis 1996, henceforth TMK).
By extending the relevant work of Mastichiadis \& Kylafis (1992,
henceforth MK92)
we allowed the inner boundary to be anything from fully reflective to
fully absorptive assigning the former case to a neutron star and the latter to
a black hole. Also, we determined the eigenvalues of
the problem as functions of the mass accretion rate  and
we found numerically and analytically the general solution 
for the spectrum as well as its asymptotic 
forms. The obtained results in the black hole case
show a remarkable
similarity with the X-ray spectra of the Galactic black hole candidates
in their high states.
In \S 2 we present the solution of the radiative
transfer problem while in \S 3 we give some examples and draw our conclusions.

\section{Radiative Transfer}
Consider a spherically symmetric accretion onto a compact
object with mass rate $\mcdot$, 
where the bulk velocity of the infalling plasma is
given by
$$
v(r) = c \left[ { {(1-\ell)r_s} \over r} \right]^{1/2} ~,
\eqno(1)
$$
with $c$ the speed of light,
$r_s$ the Schwarzschild radius and $\ell$ the ratio of the escaping
luminosity to the Eddington one. 
The Thomson optical depth of the flow is defined by
$$
\tau_T(r)=\int_r^\infty drn_e(r)\sigma_T=
\mdot \left[ { {r_s} \over {r(1-\ell)}}
\right]^{1/2} ~,
\eqno(2)
$$
where $n_e(r)$ is the electron number density,
$\sigma_T$ is the Thomson cross section, $\mdot$=$\mcdot$/$\mcdot_E$ 
and $\mcdot_E$ is the Eddington accretion rate defined by
$$
\mcdot_E \equiv {L_E\over c^2}={4\pi GMm_p\over \sigma_Tc} ~.
\eqno(3)
$$
Here $L_E$ is the Eddington luminosity, 
$M$ is the mass of the central object, $m_p$ is the proton mass
and $G$ is the gravitational constant.

Consider next low energy photons
that find themselves in the accretion flow.  As the photons diffuse outward,
they scatter off the inflowing electrons
gaining energy on average.
At some characteristic radius, called the trapping radius
(Rees 1978; Begelman 1979), the conditions of optical depth and 
velocity of the inflowing electrons are such that
the photons are advected inwards at the same rate 
as the one with which they
diffuse outwards.  Photons that either are emitted inside
the trapping radius or find themselves there 
cannot escape easily, 
but if they do, they gain significant amount of energy.

Let us investigate next the relative importance of the bulk (converging
flow) and the thermal motion of the electrons to the mean photon energy
change per scattering.
The mean energy gain per scattering of a photon by thermal
Comptonization  $<\Delta E_{tc}>$ is proportional to $(v/c)^2$,
i.e.,
$ <\Delta E_{tc}> \approx E(4kT_e-E)/m_ec^2$.
On the other hand, the mean energy  gain $<\Delta E_{cf}>$
in the presence of a converging flow
is proportional to $v/c$ (e.g., BP).
By using equations (1) and (2) we write
$<\Delta E_{cf}>\approx 4E{{d(v/c)}/{d\tau_T}}={{4(1-\ell)}/{\mdot}}$.
Thus, we have
${{<\Delta E_{tc}>}/{<\Delta E_{cf}>}}
\approx {{(kT_e-E/4)\mdot}/{(1-\ell)m_ec^2}}< {{1}/{\delta}}$,
where
$$
\delta = { {m_ec^2} \over {kT_e} }  
{(1- \ell)\over \mdot} ~.
\eqno(4)
$$
This can be further written as $\delta=
51.1\times T_{10}^{-1}\mdot^{-1}$, with $T_{10}\equiv kT_e/(10 $ keV).
Hence, the bulk motion Comptonization dominates the thermal one
if  $\mdot T_{10} < 51$.

The dominant power-law and the high-energy cutoff of the spectrum can
be understood in the following way: As it was shown above, the fractional
increase in energy of a low-energy photon in its collision with accreting
electrons of bulk velocity $v(r)$ and temperature $T_e$ is given by
${{<\Delta E>_{incr} }/{ E}}
\approx {{4(1-\ell)}/{\mdot}}+{{4 k T_e}/{m_e c^2}}$. 
At the same time, the recoil effects cause a fractional decrease in the energy
of the photon given by
${{<\Delta E>_{decr} }/ {E}} \approx - {{E}/{m_ec^2}}$.
When $E \ll m_ec^2$, the recoil effect is negligible, resulting  in a pure
power-law spectrum.
At high energies the two effects are comparable and the turnover in the
spectrum occurs at
${ {E_c}/{m_ec^2} } \approx
{ {4(1-\ell)}/{\mdot} } + { {4kT_e}/ {m_ec^2} }$,
or $E_c/m_ec^2 \approx 4(1-\ell)\mdot^{-1}$ when $kT_e \ll m_ec^2$.

The radiation spectrum observed at infinity can be found by solving the
equation of radiative transfer for the photon occupation number $n(r,\nu)$
(Eqn. [18] of BP).
We remind the reader that
the BP equation does not take fully into account general relativistic
effects; also the associated diffusion coefficient
has been derived in a fully  self consistent manner.
Both of these issues have been recently discussed (Zannias \& Titarchuk 1996,
henceforth ZT96).
 
The two boundary conditions of the radiative transfer equation 
can be formulated in terms of the spectral energy flux 
$F$ (cf. Eqn. [21] of BP).  
The first is that the total spectral flux integrated over the photosphere 
should depend only on $E$ as $\tau_T\to 0$ or $r\to \infty$. 
The second boundary condition is that we have a boundary 
with albedo $A$ at some radius $r_b$.
The net energy flux through this surface is
$$
F(r_{b},x) =
-x^3\left({{1-A}\over{1+A}}\right){n\over2} ~,
\eqno(5)
$$  
where $x\equiv h\nu/kT_e$ and $T_e$ is the electron temperature considered to 
be constant throughout the flow. From the setup of the problem, $r_b$ is 
the smallest value that the variable
$r$ can obtain. 

If the inner boundary is fully reflective ($A=1$), all input photons escape.
This is the problem considered in MK92 
who pointed out the analogy of the fully reflective boundary to a neutron
star surface. On the other hand,  
if the inner boundary is fully absorptive ($A=0$), then we have the 
equivalent of a black-hole horizon. However, since we do not take
an infinite atmosphere but we consider instead a {\sl finite} flow extending 
up to $r_b$, our case is different from the one considered in PB and 
Colpi (1988). Relativistic effects close to the horizon of the
compact object have been ignored.

The radiative transfer equation, satisfying the above formulated 
boundary conditions is solved in TMK. This solution is extended in ZT96 
by taking into account relativistic corrections.  
The spectra are calculated numerically 
by using the iteration method to solve the boundary problem 
of the appropriate Fokker-Planck equation.
The analytic approximation found by the separation of variables method  
describes satisfactorily the emergent spectra in the whole energy 
range  up to the high energy turnover $E_c\sim m_ec^2/\mdot$.
Alternatively, as 
it is shown by TMK, this approximation  can be written
as a sum of two components: The low frequency injection
is presented by a coherent solution and the high energy tail is presented
by the fundamental mode. This mode can be presented by a convolution
of the low-frequency source spectrum and the Green function
$I(x,x_0)$ given by 
$$
I={{b}\over {2\mu x_0}}
\left({x \over {x_0}}\right)^{\alpha+3+\delta}~,
\eqno(6a)
$$ 
for $x\leq x_0$ and by 
$$
I={{b}\over {2\mu x_0}}
{{e^{-x}}\over{\Gamma(2\mu)}}
\left({x \over x_0}\right)^{-\alpha}
\int_0^\infty e^{-t}(x+t)^{\alpha+3+\delta}
t^{\alpha-1}dt~,
\eqno(6b)
$$
for $x\geq x_0 $.

Here
$ b=\alpha(\alpha+3+\delta)$, 
$\mu~=~{1\over2}[(\delta-3)^2~+~4\gamma]^{1/2}$, 
$\gamma=2\lambda^2\delta$, $\lambda$ is the first eigenvalue
of the appropriate boundary problem 
and the spectral index
is determined by
$$\alpha=\mu-(3+\delta)/2.\eqno(7)$$ 
The normalization of $I(x,x_0)$ is chosen in such a way as to keep the
photon number equal to $1/x_0$. 

As we noted earlier,
the parameter which shows the importance of the bulk Comptonization 
effects relative to thermal Comptonization is $\delta$, given by 
Eqn. (4).
In the case where $kT_e\ll m_ec^2$ and $\mdot \simeq 1$, then $\delta\gg 1$
and this leads immediately to an {\it asymptotic} relation for 
the energy spectral index $\alpha$ 
$$
\alpha=2\lambda^2-3 ~.
\eqno(8)
$$ 
As long as $\delta\gg 1$, bulk Comptonization effects dominate the thermal 
ones and the value of $\alpha$ is independent of the temperature of 
the electrons.

This boundary condition, along with the radiative equation implies 
that the eigenvalues are the roots of the equation (TMK)
$$ 
\left({5\over2}-{{\hat\varepsilon}\over3}\tau_b\right)
\Phi(-\lambda^2+{5\over 2},{7 \over 2},\tau_b)
+ {{5-2\lambda^2}\over7}\tau_b\Phi(-\lambda^2+{7\over 2},{9\over 2},\tau_b)=0
\eqno(9)
$$
where $\tau_b=3\mdot/2$ and
$$
\hat\varepsilon =\alpha+3 -\left[{{3(1-A)}\over
{2(1+A)}}\right]\left[{{r_b}\over{r_s(1-\ell)}}\right]^{1/2} ~.
\eqno(10)
$$
By using Eqns. (9-10) it can be demonstrated (TMK)
  that the spectrum becomes harder
as $A$ increases. The dependence of the spectral index 
on the albedo $A$ is displayed in Figure 1.  We point out that for
$A=0$ the power-law index $\alpha$ is always larger than 1 and approaches
1.5 for $\mdot~\gg~1$.  
For the fully reflective inner boundary
case ($A=1$), $\alpha \to 0$ for $\mdot~\gg 1$. 
Thus, the hardest spectral slope ($\alpha=0$) corresponds
to the fully reflective inner boundary as it was shown in MK92.
\begin{figure}[t]
\epsfxsize=6.5 cm
\vspace{-2.5 cm}
\epsffile{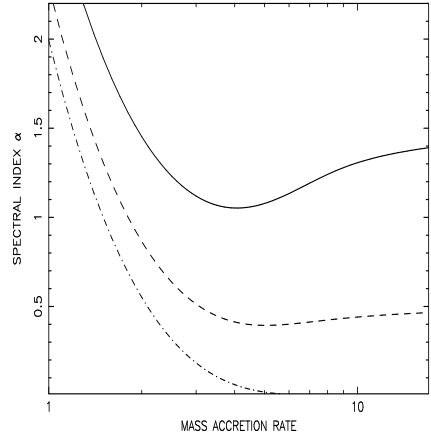}
\caption{
Plot of the energy spectral index $\alpha$ versus 
$\mdot$ for different albedo values A. 
Here $kT_e=1$ keV, $\ell=0$ and $A=0$ (solid line), $A=0.5$ 
(dashed line), $A=1$ (dash-dotted line).}
\vspace{-.3 cm}
\end{figure}

Of special interest is the case of a fully absorptive boundary $A=0$.
The obtained in this case spectral energy power-law index $\alpha$ 
is shown in Figure 2 as a function of $\mdot$. 
As it can be seen, the spectral index is a weak 
function of the electron temperature 
in the wide range of mass accretion rates $\mdot=
1-20$ and electron temperatures $kT_e=0-3$ keV.
The temperature dependence of the spectral
index becomes stronger for very high accretion rates 
$\mdot >20$ and for electron 
temperatures $kT_e>10$ keV. 

A second important consequence of the solution is that, despite the 
presence of the exponential in front of the integral 
in equation (6b), the power law can 
extend to $x\gg 1$ (see TMK). 
Thus, for $kT_e \ll m_ec^2$, the power law extends to energies 
of order $m_ec^2$ independently of the electron temperature. 
Only for energies $x>\delta$ the spectrum exhibits an exponential
turnover.

\begin{figure}[t]
\epsfxsize=6.5 cm
\vspace{-2.5 cm}
\epsffile{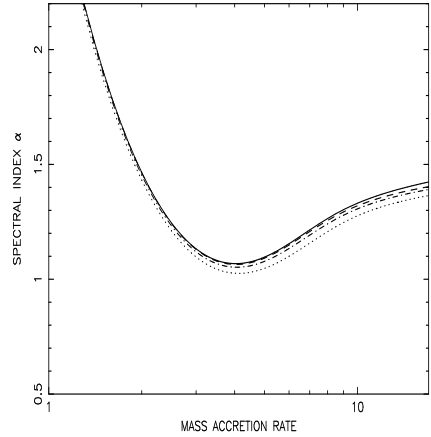}
\caption{
Plot of the energy spectral index $\alpha$ versus 
$\mdot$ for different electron temperature values $kT_e$. 
Here $A=0$, $\ell=0$ and $kT_e=0$ keV (solid line), $kT_e=0.1$ keV
(dashed line), $kT_e=1$ keV (dash-dotted line) and $kT_e=3$ keV 
(dotted line).}
\vspace{-.3 cm}
\end{figure}

Figure 3 shows a comparison between the analytical approximation
where the hard tail of the specrum is presented by Eqn. (6b)
and the numerical solution.  
The analytical approximation represents 
the emergent spectrum quite reliably in the whole energy range 
below the exponential turnover. In this region 
two effects, the bulk motion upscattering and
the Compton (recoil) downscattering,
compete forming the hard tail. Because the analytical
representation does not take into account properly the inner boundary
condition at energies $E>E_c$ there is a significant deviation of the 
analytical curve from the numerical solution close to the cut-off.
When one takes into consideration the relativistic corrections, 
the hard tail becomes flatter, since Compton
downscattering is less efficient at high energies (ZT96).

In Figure 4 the spectra are presented for different mass accretion 
rates. Here (as in Fig. 3)
we have assumed external illumination of the converging flow by the
low-energy black body radiation of an accretion disk 
having a characteristic temperature $T_{bb}$; furthermore
we have assumed a space
source distribution of the form (see, for example, Sobolev 1975)
$$
S(r) = S_0 r^{-2}\exp(-\tau_T(r)/\mu) ~,
\eqno(11)
$$
where $S_0$ is a normalization constant 
and  $\mu$ is the cosine
of the angle of incidence at $r \sim r_s$.
All spectra are pure power laws in the high energy range 
from 15 kev up to the exponential turnover, which occurs at energy 
$E_c \approx (1-\ell) m_ec^2/\mdot$.
It is worth pointing out that,
despite the fact that the accretion rate is taken to
vary from 2 to $10 \mcdot_E$, the resulting energy spectra have indices 
around 1.5. 
\vspace{-.4 cm}

\begin{figure}[t]
\epsfxsize=6.5 cm
\vspace{-2.5 cm}
\epsffile{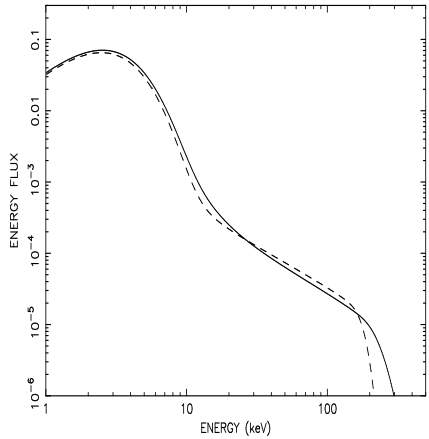}
\caption{
Plot of the emergent spectral energy flux versus photon energy
calculated by  numerical and analytical methods for $kT_e=2.5$ keV,
$\mdot=2$, $\ell=0$, $A=0$ and $kT_{bb}=0.833$ keV.
The solid line represents the numerically calculated spectrum
while the dashed line represents the analytic
approximation [the hard tail of the spectrum $E>15$ keV is described
by Equation (6b)].
The space source distribution is given by equation (11) with $S_0=1$ and
$\mu=0.3$.}
\vspace{-.3 cm}
\end{figure}
\begin{figure}[t]
\epsfxsize=6.5 cm
\vspace{-2.5 cm}
\epsffile{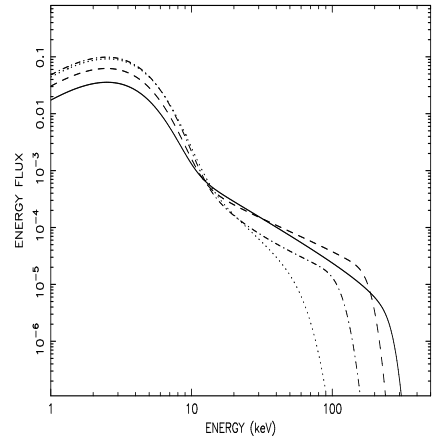}
\caption{
Analytical spectra
for different mass accretion rates
and $kT_e=2.5$ keV, $\ell=0$, $A=0$, $kT_{bb}=0.833$ keV.
The $\protect\mdot$ values are 2 (solid line),
3  (dashed line), 5 (dot-dashed line) and 10 (dotted line).
The space source distribution is given by equation (11) with $S_0=1$ and
$\mu=0.3$.}
\vspace{-.3 cm}
\end{figure}

\section{Conclusions}
We have solved  numerically and 
analytically the radiative transfer equation in the 
diffusion approximation
in the case of thermal electrons infalling spherically onto a compact
object. We have considered a finite medium (as opposed to
an infinite medium considered in the related work of PB and Colpi [1988])
taking as inner boundary a totally absorptive or reflective surface, 
in the case where the compact object is a black hole or a neutron  
star, respectively.
We have shown that the escaping spectral energy flux exhibits an extended 
power law at high energies up to a few hundred keV
and has an index that depends on the inner boundary condition.
The important parameter (in addition to the albedo introduced through
the inner boundary condition)
which determines the spectral slope is $\delta$
(eq. 4) and as long as this remains much greater than unity the upscattering
is mainly due to the bulk motion of the electrons.
Therefore we find that, for a wide range of values 
of the dimensionless accretion rate $\mdot$ and of 
the electron temperature $T_e$, the spectral slope is insensitive
to both of these quantities.

The above conclusion is especially true
in the case of accretion into a black hole (absorptive boundary).
The dominant energy
spectral index takes values around 1.5 for a wide variety of 
$\mdot$ ranging from 1.5 to 10 (see Figs. 2-4).
Furthermore, this spectral index appears to be independent of the electron
temperature as long as this remains below 10 keV (see Fig. 2).
{\it Therefore we propose that our calculations may
offer an explanation for the spectra observed by black-hole candidates
in their high states} (e.g. Ebisawa et al. 1993, Sunyaev et al. 1994
-- for a recent review see Tanaka \& Lewin 1995).
The situation is changed only when either the temperature
of the electrons is higher than 10 keV or when the escaping luminosity
approaches the Eddington limit.
In this last case the parameter $\delta$
is reduced due to the radiative force. In both cases
the spectra are dominated by thermal Comptonization
and are related to the hard spectra of black holes and neutron stars.

{\sl Acknowledgments:} L.T. 
would like to acknowledge support from NRC grant,  NASA grant
NCC5-52 and from Sonderforschungsbereich 328.
L.T. and A.M. thank the staff of the Foundation for Research and 
Technology-Hellas for their hospitality.
Also L.T. thanks Sandip Chakrabarti, 
John Contopoulos, Demos Kazanas and Thomas Zannias for extensive discussions. 
This work was partially supported by the Deutsche Forschungsgemeinschaft under
Sonderforschungsbereich 328.

{}
\end{document}